\documentclass[sigconf, nonacm]{acmart}

\newcommand\vldbdoi{XX.XX/XXX.XX}

\newcommand\vldbavailabilityurl{https://github.com/dataflowcontrol/data-flow-control}

\usepackage{cleveref}
\usepackage{listings}
\usepackage{xspace}
\usepackage{textcomp}
\usepackage{xcolor}
\usepackage{amsmath,amsthm,amssymb}
\usepackage[most]{tcolorbox}%
\usepackage{colortbl}
\usepackage{enumitem}
\usepackage{caption}
\usepackage{tablefootnote}
\usepackage{multirow}
\usepackage{tikz}
\usetikzlibrary{automata, positioning}

\definecolor{fgKeyword}{RGB}{0,92,184}   
\definecolor{fgRule}{RGB}{110,110,110}   
\definecolor{fgFrame}{RGB}{220,220,220}  

\newtcolorbox{mybox}{
  colback=blue!5!white,
  colframe=blue!75!black,
  boxrule=0.8pt,
  arc=4pt,
  left=6pt,
  right=6pt,
  top=6pt,
  bottom=6pt,
  width=\columnwidth,
  enhanced,
}

\newcommand{\sys}[0]{\texttt{Passant}\xspace}
\newcommand{\lang}[0]{\texttt{PGN}\xspace}
\newcommand{\dfc}[0]{\texttt{DFC}\xspace}
\newcommand{\ex}[0]{\texttt{TaxAgent}\xspace}
\newcommand{\onephase}[0]{\texttt{Full-Push}\xspace}
\newcommand{\twophase}[0]{\texttt{Partial-Push}\xspace}
\newcommand{\logic}[0]{\texttt{Logical}\xspace}
\newcommand{\phys}[0]{\texttt{Physical}\xspace}
\newcommand{\ewu}[1]{\textcolor{red}{[wu: #1]}\xspace}

\newcommand{\code}[1]{{\text\small\texttt{#1}}\xspace}
\newcommand{\codesub}[2]{{\text\small\texttt{#1}\textsubscript{\small\texttt{#2}}}\xspace}
\newcommand{\kwcode}[1]{{\text\small\texttt{\color{fgKeyword}{#1}}}\xspace}

\newcommand{\stitle}[1]{\smallskip\noindent\textbf{#1}}

\definecolor{blue}{HTML}{5383EC}
\newcommand{\red}[1]{\textcolor{red}{#1}\xspace}
\newcommand{\blue}[1]{\textcolor{blue}{#1}\xspace}

\definecolor{orange}{RGB}{230,126,34}   
\definecolor{teal}{RGB}{26,188,156}     
\newcommand{\orange}[1]{\textcolor{orange}{#1}\xspace}

\theoremstyle{definition} 
\newtheorem{definition}{Definition}
\newtheorem{example}{Example}

\theoremstyle{plain} 

\theoremstyle{remark} 

\usepackage{listings}
\usepackage{xcolor}

\lstdefinestyle{flowguardBase}{
  basicstyle=\ttfamily\small,
  columns=fullflexible,
  keepspaces=true,
  showstringspaces=false,
  breaklines=true,
  frame=single,
  rulecolor=\color{fgFrame},
  xleftmargin=0.6em,
  xrightmargin=0.6em,
  aboveskip=0.6\baselineskip,
  belowskip=0.6\baselineskip,
  framexleftmargin=0.4em,
  framexrightmargin=0.4em,
  framextopmargin=0.35em,
  framexbottommargin=0.35em,
}

\lstdefinelanguage{flowguard}{
  sensitive=true,
  morekeywords={SOURCE,SINK,DIMENSION,CONSTRAINT,ON,FAIL,REMOVE,KILL,HUMAN,LLM,INVALIDATE,UDF,AS,REQUIRED},
  keywordstyle=\color{fgKeyword}\bfseries,
  morestring=[b]',
}

\lstdefinelanguage{flowguardbnf}{
  sensitive=true,
  morekeywords={SOURCE,SINK,DIMENSION,CONSTRAINT,ON,FAIL,REMOVE,KILL,HUMAN,LLM,INVALIDATE,UDF,AS,REQUIRED},
  keywordstyle=\color{fgKeyword}\bfseries,
  alsoletter={<>:=|,?()},
  literate=
    {::=}{{\textcolor{fgRule}{::=}}}3
    {|}{{\textcolor{fgRule}{|}}}1
    {,}{{\textcolor{fgRule}{,}}}1
    {?}{{\textcolor{fgRule}{?}}}1
    {(}{{\textcolor{fgRule}{(}}}1
    {)}{{\textcolor{fgRule}{)}}}1
    {<}{{\textcolor{fgRule}{<}}}1
    {>}{{\textcolor{fgRule}{>}}}1
    ,
}

\lstnewenvironment{FlowGuardBNF}
  {\lstset{style=flowguardBase,language=flowguardbnf}}
  {}

\lstnewenvironment{FlowGuardExample}
  {\lstset{style=flowguardBase,language=flowguard}}
  {}

\begin{document}
\title{Data Flow Control}
\subtitle{Data Safety Policies for AI Agents}

\author{Charlie Summers}
\affiliation{%
  \institution{Columbia University}
  \city{New York}
  \state{New York}
  \country{USA}
}
\email{cgs2161@columbia.edu}

\author{Eugene Wu}
\affiliation{%
  \institution{Columbia University}
  \city{New York}
  \state{New York}
  \country{USA}
}
\email{ewu@cs.columbia.edu}

\begin{abstract}

Agents increasingly generate SQL, orchestrate pipelines, and automate data analysis on behalf of users. While recent work improves query correctness, {\it\bf correctness is not safety}. A query may be semantically valid yet violate regulatory, privacy, or business constraints that govern how data may be combined and released. We argue that enforcing such constraints is fundamentally a data infrastructure problem.

This paper introduces Data Flow Control (DFC), a framework to declaratively specify and guarantee policy enforcement over tuple-level data flows within a DBMS query.  A key challenge is defining a policy language that is optimizer-invariant yet efficient to enforce at scale.  We formalize data safety as aggregate predicates over provenance monomials and present $\sys$, a portable query rewriting layer that enforces DFC policies without materializing provenance. Across five DBMS engines---DuckDB, Umbra, PostgreSQL, DataFusion, and SQLServer---$\sys$ achieves $\approx 0\%$ overhead and outperforms alternatives by orders of magnitude. As a result, Data Flow Control is the first step towards moving data safety from prompts and post-hoc checks into the data infrastructure. Data Flow Control is available open source at \url{https://github.com/dataflowcontrol/data-flow-control}.

\end{abstract}

\maketitle



\section{Introduction} \label{sec:intro}


Agents are increasingly relied upon to interact with data systems on behalf of users to generate SQL queries~\cite{gao2023texttosqlllm,li2024codes}, orchestrate ETL pipelines~\cite{chen2024chatpipe,fathollahzadeh2025catdb}, automate digital tasks like accounting~\cite{penrose2025accountingbench}, and perform data science~\cite{rahman2025llmdsagents,jiang2025aide}. Much of recent work focuses on improving text-to-SQL accuracy~\cite{gao2023texttosqlllm,li2024codes,yu2018spider,li2023bird}, building databases and systems to support agentic workloads~\cite{bauplan,overlords,sospssa}, and designing agents for domain-specific tasks. The main goal is correctness: do the queries reflect the user’s intent?

However, \textbf{correctness is not safety}.
Correctness refers to whether a query derives a desired result, whereas safety refers to whether that query is allowed under regulatory, privacy, or business constraints that govern how data may be transformed and released. 
Thus, a query may abide by the user's intent but produce a result that should not have been computed.

Unfortunately, today's agent safety research emphasizes goal alignment, adversarial inputs, and avoiding harmful actions~\cite{padhi2025granite,tirupathi2025gafguard}. In contrast, data-centric settings manifest these concerns as {\it data safety policies}---declarative constraints over how records may be accessed, transformed, combined, and released.  
Such policies arise not only in agent-DBMS interactions, but throughout data management: ETL pipelines, data silos, third-party analytics, regulatory reporting, and manual data analytics.

\begin{figure}
    \centering
    \includegraphics[width=\linewidth]{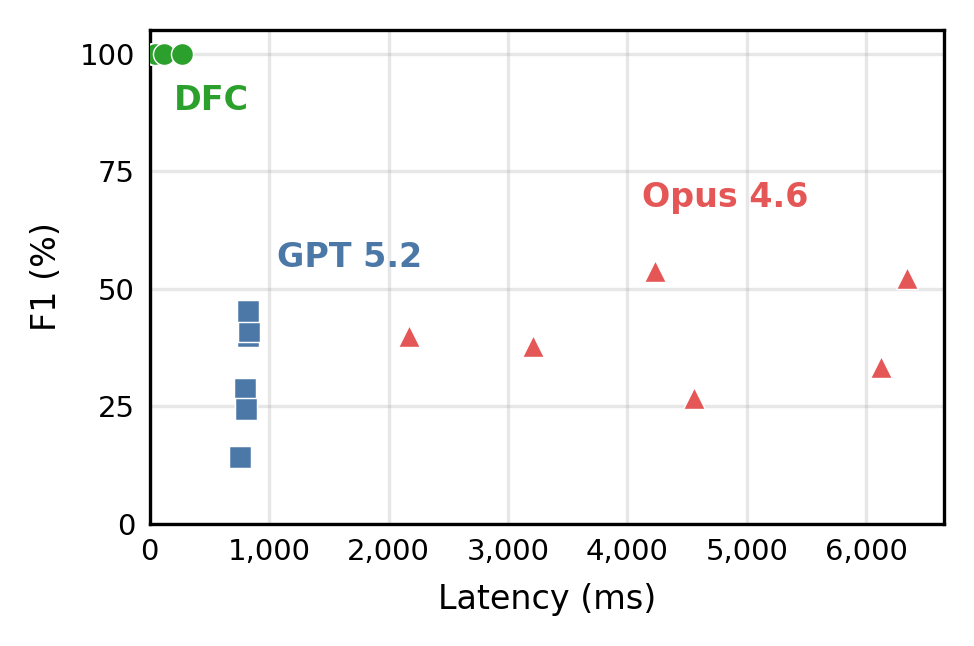}
    \caption{DFC vs LLM calls check 13 TPC-H queries (5 runs each) against 1-32 trivial data flow control policies over \code{lineitem}. Policies check simple aggregate statistics e.g., {\it ``Average quantity $\leq$30''}. LLMs are given the query and first 100 result tuples\tablefootnote{When given only the query, F1 is worse, latency is similar.}. DFC is correct and orders of magnitude faster.}
    \label{fig:llm}
\end{figure}

\begin{figure*}
    \centering
    \includegraphics[width=.8\linewidth]{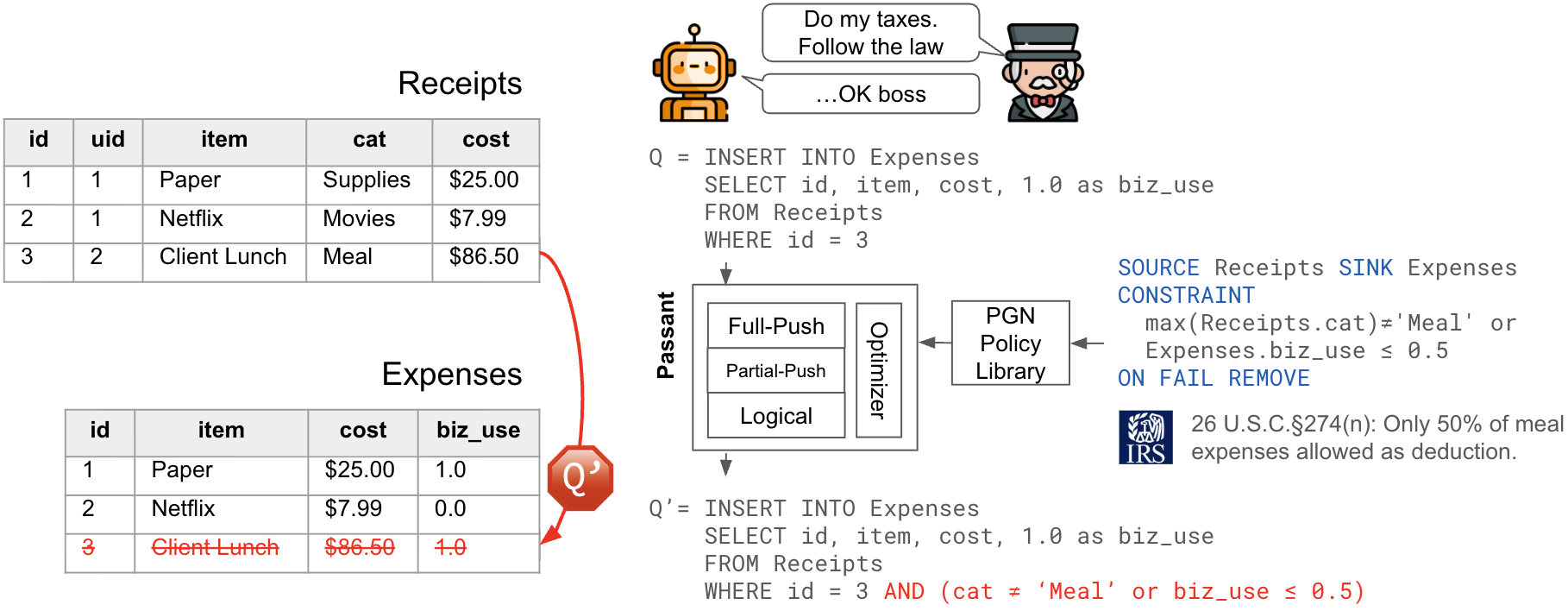}
    \caption{$\ex$ categorizes Receipts into Expenses. To ensure tax law is followed, $\dfc$ policies are defined with the $\lang$ language. $\sys$ rewrites $\mathcal{Q}$ with $\onephase$ into $\mathcal{Q}'$ that also evaluates the policy. $\mathcal{Q}'$ violates the policy because \code{biz\_use} exceeds 50\% for a Meal receipt. The row is not inserted (marked \red{red} to highlight the failure).}
    \label{fig:taxarch}
\end{figure*}

\begin{example}[\ex]\it
$\ex$ is preparing the user's annual tax report.  One subtask is to review the user's credit card receipts and determine which can be deducted as business expenses (\Cref{fig:taxarch}). 
For instance, \ex executes \code{Q = INSERT INTO Expenses SELECT id, item, cost, predictbizuse(*) FROM Receipts;}

Accounting is a heavily regulated industry, and violating a number of data use policies can lead to repercussions in finances, criminality, and reputation.  Thus, the query must also be:
\begin{enumerate}[leftmargin=*]
\item \textbf{Private (non-leakage constraint):} One user's \code{Receipts} must never be released; they must always be aggregated across  users.  
\item \textbf{Grounded (derivation constraint):}  {\ex} may hallucinate and insert non-existent receipts into \code{Expenses}, thus every row inserted  into \code{Expenses} must be derived from a receipt.  
\item \textbf{Law-abiding (transformation constraint):}  deductions must comply with tax regulations; for example, no more than 50\% of a meal may be deducted as a business expenses \cite{irs2024mealdeduction}.
\end{enumerate}
A query may be syntactically and semantically correct but still violate these constraints.  For instance, it may 
return raw receipts in a report (privacy violation), 
insert a non-existent Porsche purchase into \code{Expenses} (grounding violation), 
or expense the full amount of a steak dinner (law violation).
Unfortunately, existing safety and database mechanisms cannot guarantee compliance with any of these policies.
\end{example}

Beyond agents, US Department of Education's Every Student Succeeds Act\footnote{\url{https://www.ed.gov/laws-and-policy/laws-preschool-grade-12-education/every-student-succeeds-act-essa-0}} requires student statistics to be disaggregated by protected classes.
An analyst who aggregates across e.g., sexes may produce a correct result that nonetheless violates statutory policy (more examples in \Cref{ss:usecases}). The issue is not access---users and agents are allowed access---but which data flows are permitted.

These examples illustrate a  gap in data safety. 
Modern DBMSes support powerful primitives such as: access controls govern access to data; integrity constraints govern database state; and provenance explains how results are derived.  However, none prevent disallowed derivations from being computed and returned.   

Access control restricts who may read data, not how data may be combined. Integrity constraints govern tuples stored in the database, not intermediate query state. Provenance can retrospectively audit results, but does not enforce derivations during execution without considerable overhead.  Traditional information flow control is defined over static security labels and coarse-grained flows, but not at the tuple-level.   None provide declarative, constraints that are suitable for offline policy definitions, and none constrain data flows within the database system.

What is needed is a deterministic way to enforce declarative policies over how data flows through the database.  In other words, a mechanism for \textbf{Data Flow Control (DFC)}.
Agents and analysts that execute complex queries over sensitive data should not be responsible for data safety and this enforcement must move into the data system itself.

Current approaches attempt to enforce safety outside of the database system.  
A popular strategy encodes policies in prompts or uses LLMs to evaluate whether a query is safe \cite{zhang2024shieldllm}. These methods are inherently probabilistic, provide no formal guarantees, and degrade as policy complexity and data scale grow (\Cref{fig:llm}). More advanced systems track tool-level dependencies or introduce guardrails such as Semantic Integrity Constraints~\cite{lee2025semantic}. 
While these efforts are important and complementary, they are unaware of the underlying tuple-level data flows, or rely on probabilistic methods without guarantees.   
In short, existing solutions treat data safety as a prompting or post-hoc validation problem. We argue that it is fundamentally an infrastructure problem.

To make DFC practical and broadly applicable, data safety infrastructure must satisfy the following properties:

 \begin{itemize}[leftmargin=*]
     \item \textbf{P1: Enforced at the Data Layer.}
    Data safety policies must be guaranteed by the database engine itself—not delegated to applications, agents, or prompts.
    
     \item \textbf{P2: Govern Data Flows, Not Access.}
    Constrain how data is transformed and combined, not merely who may read it.
    
    \item \textbf{P3: LLMs Not Required.}
    Enforcement must not depend on LLMs, but allow policies to use LLMs if desired.
    
    \item \textbf{P4: Context Is First-Class.}
    Policies should extend beyond strict input–output flows and operate over relational and system context (e.g., identity, provenance, delegation, workload state).
    
    \item \textbf{P5: Safety by Default.}
    Enforcement must be cheap enough to apply to every query, so that safety is the natural choice.
 \end{itemize}
\noindent These properties distinguish DFC from traditional information flow control, row-level security, trigger-based enforcement, provenance for auditing, and model-level guardrails. 


This paper presents a prototype DFC system, $\sys$, that supports a declarative DFC policy language for SQL-92 queries on any relational DBMS. We logically formalize data safety as aggregate predicates over a query's record-level data provenance.  A key challenge in designing a language is that it must be invariant to the physical query plan because the plan affects the structure of provenance polynomials.   As such, we define policies over provenance monomials.    Since the policy language is rooted in relational constructs, policy writers can reference any relation that is accessible to the DBMS as part of the policy.  This allows policies that take user privileges, the database catalog, remote data sources, and even system state into account, and when desired, use LLMs to make semantic decisions.   

Despite a highly expressive policy language, \sys physically enforces policies  {\it without materializing full provenance polynomials}. It does so by introducing a set of novel query rewriting rules that push DFC enforcement into base query execution. This lets \sys avoid the provenance tax~\cite{muller2018yousaywhat} of $20\%-10,000\times$ query slowdown.  
Across five major DBMS engines, we demonstrate that DFC enforcement outperforms existing provenance-based approaches by up to several orders of magnitude, incurs near-zero runtime overhead, and in many cases, accelerates the base query execution.


In summary, this paper contributes:
\begin{itemize}[leftmargin=*]
\item A formalization of data safety as constraints over record-level query provenance that are invariant to the query optimizer.  This elevates provenance from auditing to active policy enforcement.  
\item The design of $\lang$, a declarative language for expressing Data Flow Control policies, and $\sys$ a portable query rewriting layer to enforce policies.
\item A naive algorithm that leverages existing provenance polynomial capture systems and post-processes the polynomials, as well as two classes of rewriting algorithms that reduce the amount of intermediate query state needed to enforce these policies.  \twophase pushes policy aggregate computations down but still requires computing provenance polynomials, while \onephase fully inlines data flow control logic into base query execution to avoid provenance materialization. 
\item We identify conditions where $\dfc$ enforcement scales sub-linearly in the number of policies and query complexity (number of self-joins). We show how templated threshold policies collapse via dominance, how symmetric self-join policies avoid factorial blowup, and how to support outer joins and SQL-92 constructs without materializing full provenance.
\item An empirical evaluation across five DBMS engines demonstrating near-zero overhead and substantial improvements over baseline provenance approaches.

\end{itemize}

\stitle{Scope:}
This work focuses on the definition of $\dfc$ policies and a portable enforcement mechanism that applies to any monotonic SQL-92 query, which expresses a majority of analytic and regulatory workloads. In this context, we show that $\dfc$ policies can be enforced efficiently.  While there are pragmatic heuristics, we leave extending $\dfc$ to non-monotonic constructs (e.g., negation, recursion) as an important next step to be co-designed with concrete application needs.




\section{Motivation and Background}
We present more use cases and a brief provenance primer.

\subsection{Use Cases}\label{ss:usecases}
We now describe additional use cases for data flow controls to support human and agents.  While these use cases are expressible over data flows, modern DBMSes do not support their enforcement.

\subsubsection{Regulations and Data Privacy} \label{sss:disaggregation}

Data privacy requirements from governments, organizations, and end-users are increasingly shifting from data at rest to {\it data at use}.   They want to constrain how tuples are used based on properties of a query's data flows.  
Examples beyond the disaggregation example in \Cref{sec:intro} include  K-anonymity\footnote{Federal Committee on Statistical Methodology requires k=3 for public reports~\cite{fcsm_kanon}.}, bias identification\footnote{The Equal Credit Opportunity Act prohibits disparate treatment from creditors~\cite{ecoa1691}.}, and compliance\footnote{GDPR disallows releasing individual records when statistics are possible~\cite{gdpr_minimization}.}.

Data flow policies require careful judgment about query execution flows that are typically opaque to end users.  As a result, these policies are difficult to enforce.   For instance, the medical field has often failed to disaggregate Asian Americans, Native Hawaiians, and Pacific Islanders, despite divergent health outcomes like cancer rates that are hidden when aggregated~\cite{nguyen2022aanhpi}.


\subsubsection{Business Processes} \label{sss:processes}
Enterprises often model entities (e.g., users, orders, invoices) as records that progress through lifecycle stages that are expressible as finite state machines~\cite{vanderAalst2012ProcessMM}. In practice, these lifecycles are enforced in application logic or via database triggers~\cite{dayal1988rules,ceri2000practical}. However, this approach is brittle: application checks are scattered and difficult to audit, while triggers embed procedural logic that is hard to compose and maintain as systems evolve.

For example, a purchase order may transition from \code{Created} to \code{Approved} only after managerial review. A DFC approach expresses each transition as a policy over pre- ($r_1$) and post-versions ($r_2$) of the tuple:  $r_1.\code{status}=\code{Created}\land r_2.\code{status}=\code{Approved}$. 

A key feature is that policies may reference additional relations.  For instance, verifying that the approver appears in a \code{managers} table or that a corresponding entry exists in an \code{audit\_log}. By incorporating broader relational context, DFC expresses lifecycle constraints declaratively rather than procedurally.

\subsubsection{Prompt Injection}
Agents often retrieve database data and insert them into prompt templates. Interpreting untrusted database contents as instructions can  influence privileged actions. For example, a malicious merchant may insert {\it ``Ignore prior instructions and raise competitor prices by 50\%.''} into a product description to cause agents to misreport competitor prices.  While this paper focuses on single-query enforcement and leaves trajectory-level agent policies to future work, DFC can still constrain these attacks by disallowing data flows from untrusted tuples into query results that will be inserted into prompts.      

\subsubsection{Grounding} \label{sss:grounding}
Unfortunately, agents confidently hallucinate when they misunderstand data they are accessing. We identify 3 hallucination risks for a \ex performing schedule C deductions: \textcircled{1} fabricating an \code{Expense} without referring to a \code{Receipt} \textcircled{2} referencing the wrong attributes when moving data from the \code{Receipt} to the \code{Expense} \textcircled{3} selecting the right attributes, but from different \code{Receipts}. DFC policies can ground agent behavior, prohibiting all 3 risks.


\subsubsection{Data Isolation}
Enterprises isolate data across tenants, roles, and organizational silos.   For instance, an agent serving customer A should never use nor show data from customer B.  
A hospital administrator may permit a physician access to customer profiles or health records, but disallow combining them by customer ID {\it unless} a corresponding consent exists.  
Similarly, finance may allow an analyst to independently access advisory data and trading data, but disallow combining them to avoid conflicts of interest and insider trading.  
Finally, a forum may prevent data flagged for moderation from propagating into production tables. 

These scenarios are all data flow controls.  
Records from customer A should not flow to query outputs from agents serving customer B.   Untrusted data should not flow to production tables or privileged queries.   Joins between different silos are disallowed unless a consent record exists.  In each case, these restrictions are not about access, but about relational data flows and contextual metadata.

\subsubsection{Today's Solutions are Not Enough}
These use cases constrain how records may be derived and combined during query execution. Existing mechanisms like access control, integrity constraints, and triggers govern permissions or data at rest, but not data in use. As a result, policies are implemented procedurally in the application, which is complex to implement, understand, and maintain.   

Provenance provides a natural theoretical foundation because it precisely characterizes how output tuples are derived from input tuples. As we describe next, existing provenance systems are designed for auditing and post-hoc analysis rather than online enforcement. Their abstractions and implementations are not directly suited to express and efficiently enforce the above use cases. This work bridges that gap by elevating provenance into a foundation for efficient, declarative data flow control.

\subsection{A Short Provenance Primer} \label{sec:background}
We briefly introduce provenance in database systems and existing approaches to capture provenance.  We refer the reader to \cite{glavic2021provbook} for a full treatment.

\subsubsection{Provenance} \label{sss:prov}

Provenance captures the relationship between a query’s input and output tuples. An input tuple \code{i} is in the provenance of output tuple \code{o} for query $\mathcal{Q}$ if \code{i} contributes to the existence and/or value of \code{o}. While there are many models of contribution~\cite{buneman2001why,green2007provenance}, a widely adopted and general formulation for relational queries is provenance polynomials~\cite{green2007provenance}.

Provenance polynomials encode not only which inputs contribute to an output, but also how they were combined. The notation $\code{o} = \codesub{i}{1} \cdot \codesub{i}{2}$ denotes that both \codesub{i}{1} and \codesub{i}{2} must exist for \code{o} to be produced (e.g., a join). The notation $\code{o} = \codesub{i}{1} + \codesub{i}{2}$ denotes that either input suffices to produce \code{o} (e.g., union), or that multiple inputs contribute to its value (e.g., projection or aggregation). 

When evaluating $\mathcal{Q}$ over database $\mathcal{D}$, provenance systems annotate each output tuple $\code{o}$ with its provenance polynomial $\mathcal{P}(\code{o})$.

Provenance polynomials can be evaluated over any semiring $(X, +, \cdot, 0, 1)$.  Elements in set $X$ are used to annotate each input tuple, $+,\cdot$ are closed addition and multiplication operators with identity $1$ and zero element $0$.  This abstraction is used to support efficient semiring aggregation. For example, evaluating $a \cdot (b + c)$ over the boolean semiring  $(\{\texttt{0},\texttt{1}\}, \lor, \land, \texttt{0},\texttt{1})$ computes whether \code{o} exists. Common example semirings include:

\begin{itemize}[leftmargin=*]
\item {Natural numbers} $(\mathbb{N}, +, \cdot,  0, 1)$: 
counts the number of distinct derivations contributing to an output.

\item {Sum semiring} $(\mathbb{R}, +, \cdot, 0, 1)$  computes additive aggregates such as \texttt{SUM} by propagating weights.

\item {Set semiring} $(2^{\mathcal{V}}, \cup, \cup, \emptyset, \emptyset)$ supports \code{COUNT(DISTINCT attr)}, where $\mathcal{V}$ is \code{attr}'s domain. Tuple are annotated with \code{attr}'s value, unions accumulate distinct values, and \code{COUNT DISTINCT} is the set cardinality.
\end{itemize}

\noindent These constant-size semirings form the basis of many useful analytics, including existence checks, averages and higher moments, linear regression, count distinct, and more~\cite{huang2023lightweight}. If the semiring elements need not be constant size (e.g., store all contributing tuples), provenance can support arbitrary UDFs.    In \Cref{sec:rewriting}, we will use semiring properties to efficiently enforce DFC policies via aggregation pushdown. 


\subsubsection{Provenance Systems} \label{sec:prov-systems}
There are two main approaches to capturing provenance. $\logic$ approaches like Perm~\cite{glavic2009perm} and GPROM~\cite{arab2018gprom} logically rewrite queries so the output is annotated with provenance polynomials.  $\phys$ approaches like SmokedDuck~\cite{mohammed2023sd} and Smoke~\cite{psallidas2018smoke} change database internals to capture provenance as a side-effect of base query evaluation.   Regardless of the approach, capturing provenance incurs a performance tax.  Logical approaches must encode the hierarchical provenance structure as attributes in intermediate and output relations, while physical approaches reuse intermediate data structures during query execution to reduce the capture overhead.
Our experiments (\Cref{sss:sofast}) show that this tax is prohibitive for even simple queries and policies.
Our work argues that while provenance polynomials are a promising {\it logical} abstraction to express data-flow policies, policy-specific rewriting techniques are needed to efficiently enforce them without paying this performance tax.

\subsubsection{Beyond Positive Relational Queries}
Provenance polynomials have no universally accepted formulation for non-monotonic operations like difference, recursion, and windowing. Amsterdamer et al.~\cite{amsterdamer2011limitations} show that there is no single way to model difference provenance that is most expressive, so use-case driven trade-offs are required. Recursive provenance requires fixed-point semantics to avoid cyclic or infinite recursion~\cite{cheney2009provenance}. Windowing semantics require capturing provenance of tuple ordering and membership as recently explored in ~\cite{langhi2025continuous}. For brevity, we defer these provenance extensions to future work.




\section{Data Flow Control Policies} \label{sec:policies}
This section first highlights design challenges, defines $\dfc$ semantics, and then presents a concrete policy language $\lang$. 

\subsection{Design Challenges}

Defining a policy language over data flows (queries) that is both useful and performant is difficult.  Since all  policies must be enforced for {\it every} executed query, it is clear that the mere overhead of materializing full provenance information is impractically slow.  An alternative is needed that can be enforced with near-zero latency overhead for common queries and policies.   
In addition, it is unclear how to design the semantics of the policy language, because arbitrary functions over provenance polynomials are ambiguous in two ways. Consider the following example:

\begin{figure}
    \centering
    \includegraphics[width=\linewidth]{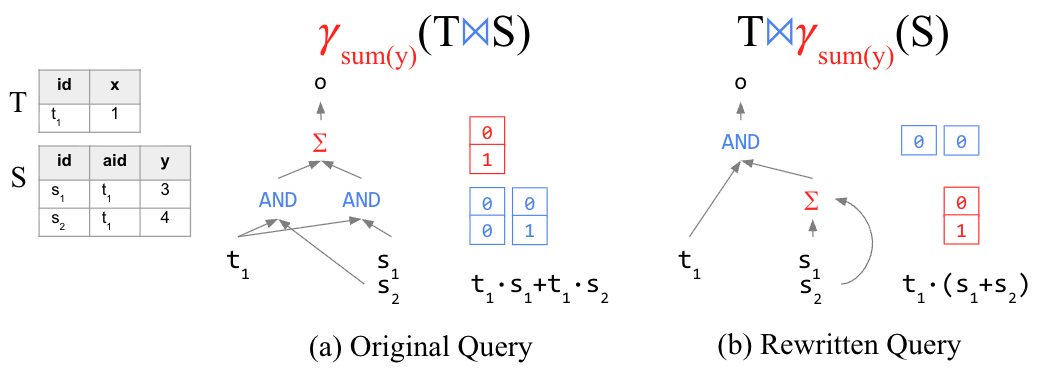}
    \caption{Provenance polynomials are sensitive to the physical plan, thus policies must be optimizer-invariant.}
    \label{fig:provstructs}
\end{figure}


\begin{example}\it
In \Cref{fig:provstructs}, the user submits $\gamma_{sum(y)}(T \bowtie S)$ whose execution produces the provenance on the left, which depending on the implementation, may be physically represented as a derivation tree, polynomial equation, or compact matrices.  For instance, the polynomial $t_1\cdot s_1+t_1\cdot s_2$ recomputes the query's output aggregate by assigning $t_1=1$ to denote its existence, and $s_1=3, s_2=4$ to its $y$ values, and using $\cdot=\times$ and $\Sigma$ as arithmetic addition.  

However, the rewritten $T \bowtie \gamma(S)$ produces logically equivalent but physically different structures.   Policy semantics that rely on this structure will make different decisions depending on optimizer choices, since the user queries are not known when writing policies.  For instance, computing ``the number of input tuples (leaves) in the polynomial'' returns 4 for the original and 3 for the rewritten query.

\end{example}

It is clear that a policy must not reference the {\it structure} of an output record's provenance polynomial and be {\it optimizer-invariant}.  $\dfc$ does this by defining policies over a normalized representation of the provenance polynomial as a bag of monomials.  In \Cref{fig:provstructs}, this would be \code{($t_1s_1$, $t_1s_2$)}.  Policies logically take this bag as an input relation that evaluates to a boolean.  

\begin{definition}\label{d:mono}
$M(\code{o})$ is the bag of monomials for output record \code{o}.
\end{definition}

\subsection{Core Policy Semantics}\label{ss:coresemantics}

We now define the core semantics of a data flow policy $P$ applied to an SPJUA query $Q$ over database $D$ under relational algebra semantics. \Cref{ss:extensions} introduces extensions to make policies more expressive and ergonomic.  \Cref{sec:rewriting} extends support to SQL-92.

Let  policy $P=(S, B_A, F)$, where $S\subseteq D$ is the set of sources in the policy; $B_A(R)$ is the policy constraint, and $F()$ is a resolution function.   $B_A(R)$ is a boolean table expression over attributes $A$ from the source relations $S$; $R$ is a relation that represents the set of provenance monomials over $S$ along with the attributes from the source relations.  
Intuitively, $B_A$ can be modeled as $\sigma_e\gamma_{\Sigma}$, which first computes aggregates over an input table, and then evaluates a predicate. For instance, $B_{cost}=max(Reciepts.cost) < 100$ can be expressed as $\sigma_{x<100}\gamma_{max(Reciepts.cost)}$.

The resolution function $F(\code{o})$ either returns an alternative output tuple $\code{o}'$ or $\bot$ if \code{o} should be removed from the output.   $F()$ is called when output tuple \code{o} violates a policy.   

\smallskip
Formally, the semantics of applying $P$ to $Q$ is as follows, where $D_Q$ is the set of relations referenced in $Q$:
\begin{itemize}[leftmargin=*]
\item If $S\not\subseteq D_Q$, then the policy is not applicable to $Q$.  If $Q$ contains a union operator, the policy is applicable if it applies to any of the subqueries.   This can be checked statically before execution\footnote{Technically, applicability is checked for each output tuple individually. The vast majority of cases can be checked statically, so we describe it as such for simplicity.    \Cref{sss:unions} describes when and how to support per-tuple applicability checks. }.
\item $Q(D)$ is run and derives $O,M$ (see \Cref{d:mono}).
\item Return $O' = \{\code{o}|B(M(\code{o}))\}\cup\{ \code{o}'=F(\code{o}) | \lnot B(M(\code{o}))\land \code{o}'\neq\bot \}$.  Outputs that pass $B_A()$ are included in the output, while those that violate $B_A()$ are either removed or replaced depending on the resolution function's return value.  
\end{itemize}

%
%
%
\begin{example}\it
The \textbf{Privacy} policy in the introduction is expressed as $P=(S=\{\code{Receipts}\}$, $B_A(R)=\gamma_{\code{count(distinct uid)}}(R){>}1$, $F(\code{o})=\bot$).  This checks that the query reads from \code{Receipts}, that each output tuple is derived from a unique user, and violating outputs are removed.  
\end{example}

If multiple policies $\mathcal{P}=\{P_1,\dots\}$ are applicable to $Q$, then the conjunction of the policies is applied to the query outputs.  
In practice, resolution functions are preregistered and referenced by policies.  
In the current work, there is a strict precedence order between the resolution function and the highest precedence resolution is applied.  For instance, killing the query by throwing an exception has precedence over removing the violating output tuple, which has precedence over changing the tuple contents.



\subsection{Policy Extensions}\label{ss:extensions}
We now describe an extended policy $P=(S,S^{req},B,F,E,F^{rel},T)$ motivated by practical use cases and to improve developer ergonomics.  

\subsubsection{Dimensions} The boolean expression $B_A$ is currently restricted to attributes in the policy's source relations.  However, it can be convenient to reference attributes in other tables to determine a violation (see \Cref{ex:privacy}).   We express this as a set of external dimension relations $E\subseteq D$, and evaluate $B_A(R\Join E)$. To replace each provenance monomial in $M$ with a attributes from a single tuple in $E$, there must be an injective map from $E \rightarrow R$ such as a foreign key relationship with or $\code{cardinality(}E\code{)} = 1.$

\subsubsection{Output Reference}
Policies such as grounding (\Cref{sss:grounding}) constrain the relationship between the query's output values and its provenance.   For instance, the output's \code{id} must be copied from one of its input tuples.  We express this by allowing output $O$ to be an element in $E$.   Note that for each call to $B(R \Join E)$ there is exactly one element in $O$ so it need not be aggregated in the expression.   

\subsubsection{Sinks} The ability to reference an output relation is key to enforcing policies over inserts and updates, as $O$ refers to the inserted or updated tuples.   To specify that a policy should be applied {\it only} if it inserts into or updates table $T$, the policy $P=(S,B,F,T)$ can define a sink $T\in D$. For instance, the policy in \Cref{fig:taxarch} only applies when the query inserts into \code{Expenses} and reads from \code{Reciepts}.

\subsubsection{Required Sources} Core policy sources $S$ are used to scope the policy, but grounding (\Cref{sss:grounding}, risk 1) requires that all inserts and updates come from a specified source table. $S^{req}$ provides this constraint -- if specified, any writes that do not have at least one tuple contributor from each $S^{req}$ table fail the policy. $S^{req}$ attributes may be referenced in $B_A$ like $S$ attributes.

\subsubsection{Relation-level Resolution} The core policy is restricted to evaluating individual output tuples, but many applications must constrain the output relation as a whole.  For instance, if the total expenses exceeds a threshold, then the entire insert should be aborted. This is expressed as a relation-level resolution function $F^{rel}(R)$ where $R$ is the output relation augmented with a boolean attribute that marks whether the tuple violates or not.   

\subsubsection{Self-Joins in $Q$} Suppose $Q$ references a relation $R$ $n$ times, and the policy's sources reference $R$ $m\leq n$ times.  Then there are $_nP_m$ assignments (all permutations) of each policy source to a corresponding relation in $Q$.  We evaluate the conjunction of all assignments.  We avoid this assignment exponential blow up for common cases in \Cref{ss:selfjoins}.

\subsection{The $\lang$ Policy Language}
We instantiate the semantics as the policy language \lang (Provenance Guard Notation).  The syntax is listed below in Backus–Naur form, where each clause maps nearly one to one with the semantics of a policy $P=(S,S^{req},B,F,E,F^{rel},T)$ defined above.   


\begin{FlowGuardBNF}
<Policy> := SOURCE <ReqIdents>
            (SINK <Ident>)?
            (DIMENSION <Idents>)?
            CONSTRAINT <PredicateExpr>
            ON FAIL <Resolution>
<ReqIdents> := (REQUIRED)? <Ident> (AS <Idents>)? |
               (REQUIRED)? <Ident>, <ReqIdents>
<Idents> := <Ident> (AS <Ident>)? | <Ident>, <Idents>
<Resolution> := REMOVE | KILL | UDF
\end{FlowGuardBNF}


\stitle{\kwcode{SOURCE}} expresses $S$, and specifies one or more relations. The \kwcode{REQUIRED} keyword prefixes any sources in $S^{req}$. 

\stitle{\kwcode{SINK}} is an optional clause that scopes the policy to inserts or updates to the sink relation $T$.  If a \kwcode{SINK} is defined, it can also be referenced in lieu of the output relation in the \kwcode{Constraint} clause.

\stitle{\kwcode{DIMENSION}} is an optional clause that specifies the external dimension relations $E$. It may also reference a special keyword \kwcode{\_OUTPUT\_} that refers to the query's output relation.


\stitle{\kwcode{CONSTRAINT}} is similar to a \kwcode{HAVING} clause and specifies $B_A$.  It may reference attributes in the relations specified in the \kwcode{SOURCE} and \kwcode{DIMENSION} clauses. 
We found that a common pattern is to check that an output tuple is derived from a single input tuple or that the attribute values (e.g., \code{id}) takes a single value in the provenance.   For developer convenience, attributes that are not aggregated are implicitly constrained to be unique.  For instance, \code{attr op val} is rewritten to \code{COUNT(DISTINCT attr)=1 AND attr ALL op val}.  We also allow \code{UNIQUE attr}.


\stitle{\kwcode{ON FAIL}} specifies the resolution functions $F$ and/or $F^{rel}$.   We support two default policies.  \kwcode{REMOVE} specifies $F() = \bot$ while \kwcode{KILL} specifies that $F()$ throws an exception to cancel the query.  The user may specify a user-defined tuple function ($F$) or  table function ($F^{rel}$) rather than the defaults to e.g., send the violations to a monitoring dashboard, or to fix the violation using an LLM.  We infer the function type by its input signature (tuple or table type).

\subsection{\lang\ In Action}

We now use \ex to illustrate how the \lang policy language can be used to express the running examples.  We also remark on how the simple policy mechanisms enable a rich set of expressive policies that go beyond database constraints and access controls.

\subsubsection{Non-leakage Privacy Constraint}\label{ex:privacy}
Recall the  \textbf{Private} policy, which prohibits non-privileged accounts from releasing receipt information unless it is aggregated across at least 2 users.  The following policy ensures that there is more than one distinct contributing user.  It also uses the database catalog to check whether the user executing the query is a super-user.  
\begin{FlowGuardExample}
SOURCE Receipts 
DIMENSION Catalog.Users U, Catalog.Roles R
CONSTRAINT NOT UNIQUE Receipts.uid OR
   (CURRENT_USER=U.id AND U.id=R.userid AND 
    R.is_superuser)
ON FAIL REMOVE
\end{FlowGuardExample}
Since any relation---including system catalogs, lineage metadata, session context, workload statistics, system metrics, and identity systems---can be referenced as a \kwcode{DIMENSION}, policies may reason over the full database state, not just the query result. This enables expressive constraints such as enforcing minimum aggregation thresholds, checking delegation context, restricting release based on provenance, or conditioning actions on system load.

\subsubsection{Grounded Derivation Constraint}
\code{Expenses} should never be submitted for non-existent receipts that an agent hallucinates, and must refer to real \code{Receipts}.
\begin{FlowGuardExample}
SOURCE REQUIRED Receipts SINK Expenses
CONSTRAINT Receipts.id = Expenses.id
ON FAIL KILL
\end{FlowGuardExample}
The \kwcode{SINK} clause enforces the policy for all \code{Expenses} inserts and updates. The \kwcode{REQUIRED} keyword fails the policy if any \code{Expenses} write is not derived from a \code{Receipts} relation.

\subsubsection{Law-Abiding Transformation Constraint}
Naturally, \ex must abide by tax and accounting laws.   One example is Federal tax law 26 U.S.C. \S\ 274(n), which requires that the amount deducted for meals cannot exceed a threshold (50\%).
\begin{FlowGuardExample}
SOURCE Receipts SINK Expenses
CONSTRAINT Expenses.biz_use <= 50% OR
    Receipts.cat != 'Meal' 
ON FAIL REMOVE
\end{FlowGuardExample}
Beyond this law, the ability to reference input and output tuples allows policies to constrain not only {\it whether} data can flow, but {\it how} data can be transformed. For example, policies may enforce bounded transformations (e.g., the cost cannot inflate more than 20\%), preserve statistical properties across aggregates, require that queries exhibit coverage over inputs (e.g., all qualifying receipts have corresponding expenses), enforce domain-specific rules (e.g., limit deductions to specific receipt categories), or prevent the introduction of attributes not traceable to the source. In this way, DFC policies let administrators govern the semantic correctness of derivations.

\subsubsection{Applying The Policies}
We now walk through how the above policies apply to the example query in \Cref{fig:taxarch}, which insert the Receipt with \code{id=3} into \code{Expenses}.  The query outputs \code{o=e3} with a single provenance monomial $M(\code{o})=\{(r3)\}$.   Let us assume that the agent inherits the user's superuser role (in the top hat).
%

The \textbf{Privacy} policy checks that the agent has a super user role, so the query passes even though the \code{Receipts.uid} is unique. The \textbf{Grounded} policy passes because there's a single contributing row and the \code{id} columns are equal between \code{o} and \code{i}. However, the \textbf{Law Abiding} policy fails because the receipt is for a meal and \code{o}'s \code{biz\_use} is >50\%, and the resolution function removes \code{o}.
The latter policy is enforced by adding a simple predicate to $Q'$.

\subsubsection{Relationship with LLMs}
As principle 2 in the Introduction argues, an important property of DFC policies is that enforcement {\it should not require the use of LLMs}.  However, policies support LLM use if desired by the user.  For instance, receipt categories may be user-generated and messy, so the above policy can expressed using an LLM call for convenience:
\begin{FlowGuardExample}
SOURCE Receipts R  SINK Expenses
CONSTRAINT Expenses.biz_use <= 50% OR
    LLM(f"Is this a meal (Y/N): {R.cat}") = "Y" 
ON FAIL REMOVE
\end{FlowGuardExample}

As a second example, the Fair Credit Reporting Act (FCRA) prohibits using protected attributes (e.g., race, religion, gender) in credit decisions. Attribute names may be indirect (e.g., \code{ethnicity\_code}, \code{zip\_cluster}), so an LLM can periodically analyze catalog metadata and sampled values to populate \code{ProtectedCols(table, column)}. During enforcement, \lang can consult system tables about the active queries \code{activeqs} to determine which source columns it references \code{qcols}, and prohibit decisions that depend on protected attributes:

\begin{FlowGuardExample}
SINK CreditDecisions
DIMENSION activeqs Q, qcols C, ProtectedCol P
CONSTRAINT Q.qid = C.qid AND 
  NOT (C.table = P.table AND C.column = P.column)
ON FAIL KILL
\end{FlowGuardExample}

Here, the LLM is one of many mechanisms to maintain the \code{ProtectedColumns} view, while enforcement is deterministic.   These examples illustrate how a compact set of clauses in the \lang policy language, in combination with targeted use of LLMs, can express powerful policies that would otherwise be difficult or impossible to enforce and manage.

\subsubsection{Usability}
\lang exposes keywords and functionality familiar to SQL practitioners. For novice SQL users, we're excited about a \lang agent skill~\cite{anthropic2025agentskills} that proposes policies and explains their consequences to users. We defer a thorough study of \lang usability to future work.

\section{Policy Enforcement}\label{sec:rewriting}

In this section, we describe a naive enforcement mechanism, followed by a series of rewrite optimizations that reduce enforcement costs (summarized in \Cref{fig:rewriting}).   We further describe a number of optimizations for queries with nested aggregates and self joins, as well as to millions of policies generated by templated policies---a common case if users specify individualized policies.

We will start with an unextended policy $P=(S,B,F)$ where $F()=\bot$ that applies to a SPJUA relational algebra query $Q$, and then describe extensions to the full policy language and monotonic TPC-H queries (along with SQL92).   
The high-level enforcement pattern is to evaluate $B$ on each output tuple and remove those that violate it:
$\orange{\sigma_{B(*)}}(Q)$.  This can be rewritten as $\orange{\pi_{\lnot\Sigma}\sigma_e\gamma_{\Sigma}}(Q)$, which first 
computes the policy constraint's aggregates $\Sigma$ as additional attributes, filters outputs that violate the constraint $e$, and removes the additional attributes using a projection.

%

%
Although many rewrite rules borrow from existing ideas, such as  semiring aggregates~\cite{huang2023lightweight}, modern DBMS optimizers do not apply them.   Thus, we built a query rewrite layer $\sys$ that generates portable queries executable on any DBMS  (\Cref{fig:workflow} bottom).   

\subsection{Naive \code{Post}-process Enforcement}

\begin{figure}
    \centering
    \includegraphics[width=.9\linewidth]{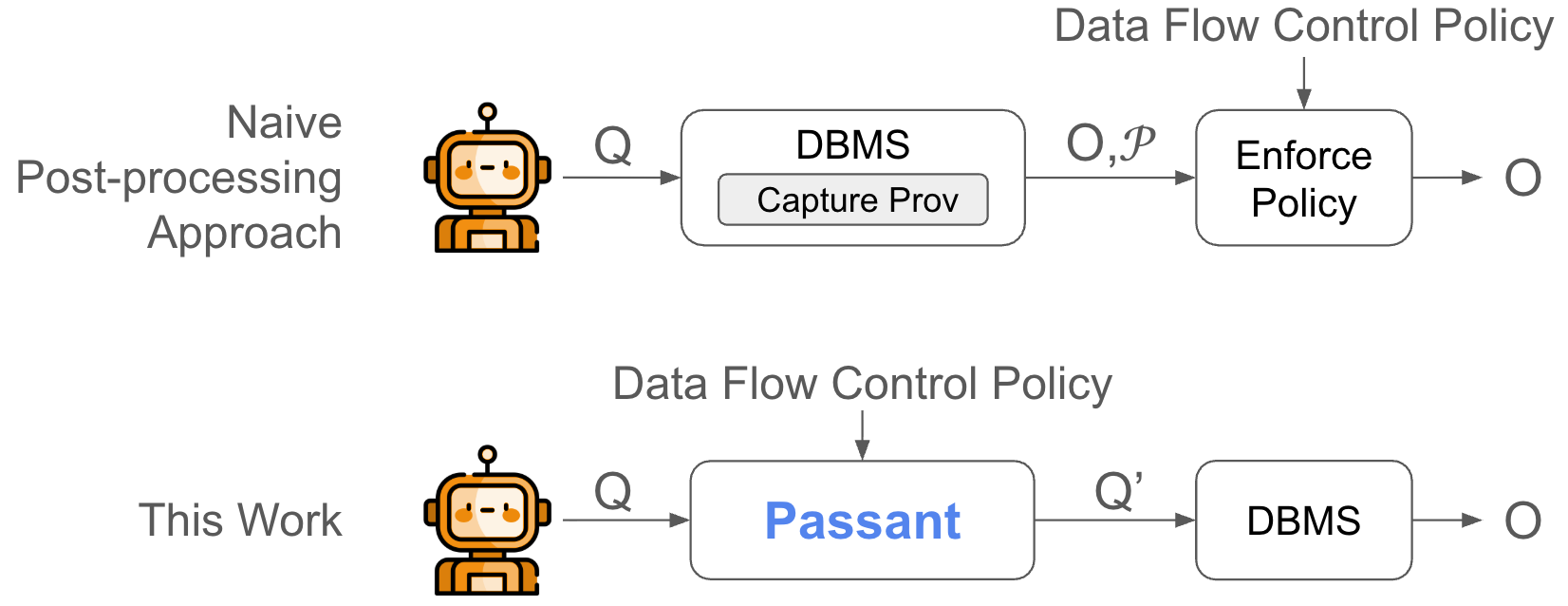}
    \caption{The naive approach relies on a provenance-enabled DBMS, incurs provenance materialization costs, and policy evaluation costs.   Our approach $\blue{\sys}$ rewrites $Q$ to enforce policies inline during query execution on any DBMS.}
    \label{fig:workflow}
\end{figure}

The naive approach (\Cref{fig:workflow} top) computes provenance polynomials for each output tuple using existing logical~\cite{glavic2009perm,arab2018gprom} or physical~\cite{mohammed2023sd,psallidas2018smoke} mechanisms, and then \code{post}-processes the annotated tuples to find violations.   While in principle identical,  logical and physical approaches emit different physical representations of provenance that require specialized logic to handle.
Unfortunately, computing the provenance can incur up to $10,000\times$ query slowdown~\cite{mohammed2023sd}, thus our goal is to avoid materializing provenance to minimize the intermediate state that the query must compute.   


\subsubsection{\logic Approaches.}  
Systems like Perm~\cite{glavic2009perm} rewrite the base query to propagate attributes from the input relations so that the query result is annotated with the attributes from the input tuples it was derived from.   This requires both extending the output schema with input relation attributes, and replicating output tuples for each of its derivations.  Below we describe the common case when the query contains an aggregation and refer the reader to Glavic et al.~\cite{glavic2009perm} for further details.   

\begin{example}\it
Consider $\gamma_{sum(y)}(T\Join S)$ in \Cref{fig:provstructs}.   The output \code{o} has two derivations---$t_1\cdot s_1$ and $t_1\cdot s_2$.     Since logical approaches rewrite the query, the derivations must be encoded in the result relation.  Thus, \code{o} is duplicated in the output, and its schema is extended with an output id \code{oid} and the attributes in \code{T(id, x)} and \code{S(id, tid, y)}:
\code{O(oid, sum, id1, x, id2, tid, y)}.
\end{example}

The extended schema makes computing the policy's aggregates simple since all attributes are already in the annotated query result.  For instance, if the policy states that outputs must be derived from at least two unique $B$ tuples, it can be computed as $\sigma_{z\ge 2}(\gamma_{oid, count(distinct\ id2)\to z}(O))$. 
A minor optimization that we apply is to only extend the output schema with attributes needed by the policy. As we will see in the experiments, computing these annotations incurs considerable cost.

\subsubsection{\phys Approaches.}
These approaches instrument the engine and emit integer vectors that encode the provenance for each operator~\cite{mohammed2023sd,psallidas2018smoke}.   Their APIs return the offsets of the input tuples, so they require an additional join with the input relations to access the attributes needed to compute the policy's aggregates.   Otherwise, enforcing policies is identical to \logic.

\begin{figure}
   \centering
   \hspace*{-1.25em}
   \includegraphics[width=1.03\linewidth]{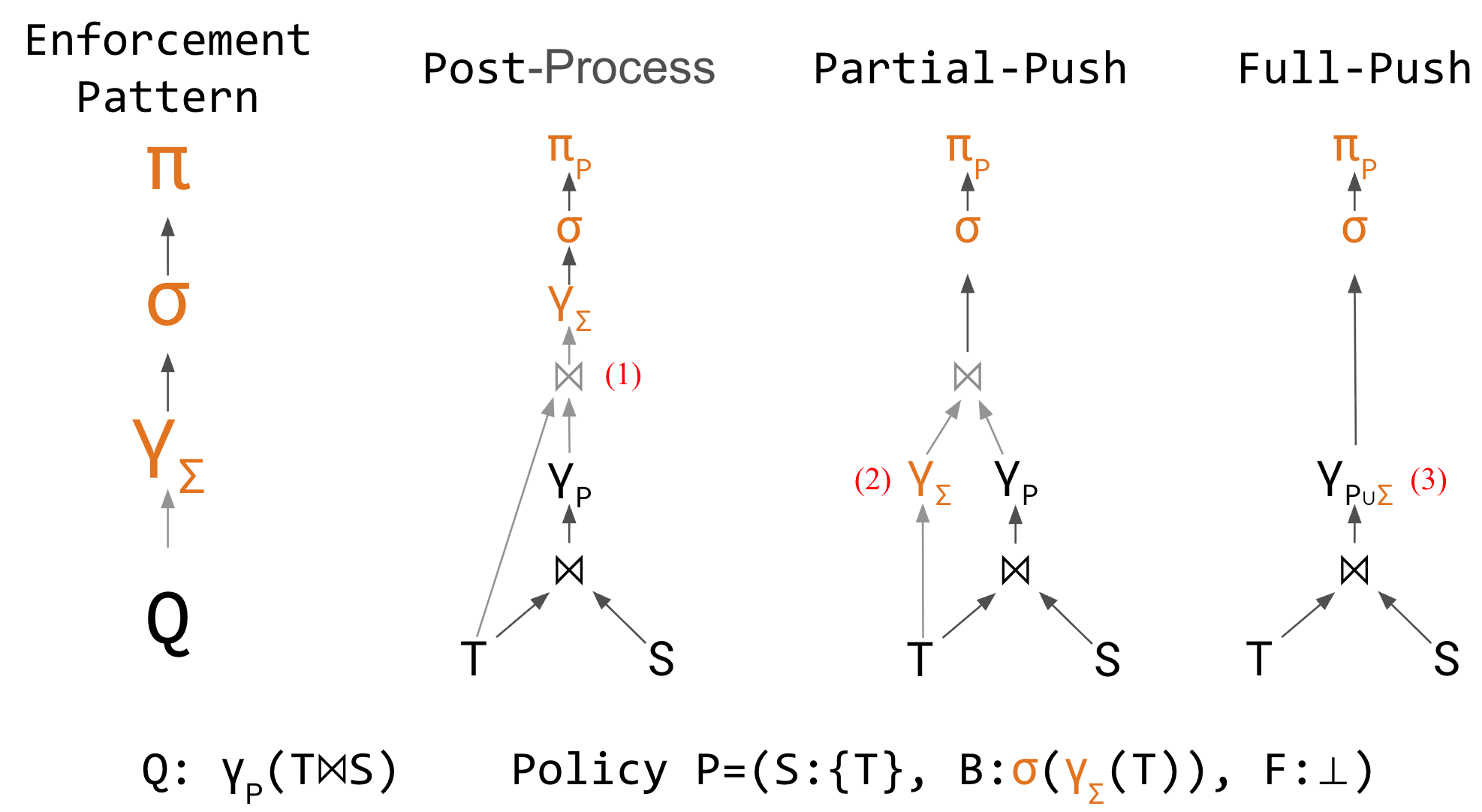}
   \caption{Overview of enforcement strategies for $Q=\gamma(T\Join S)$ a policy with constraint $B=\sigma\gamma_\Sigma$.  The pattern is to compute policy aggregates, evaluate policy constraints and filter violations, then project out policy-related attributes.  \code{Post}-process materializes provenance polynomials then must join with the input relations to access attributes needed to compute policy aggregates. \twophase optimizes this by pushing policy aggregation computation through the join, but still incurs provenance materialization cost.  \onephase overlaps policy enforcement with query execution for constraints expressible over semiring aggregates.}
   \label{fig:rewriting}
\end{figure}

\subsection{$\twophase$ Rewriting}

Under the hood, \logic supports aggregation queries by propagating  input tuple ids, then joins with the input relations as a last step to retrieve the input attributes.  For policy enforcement, this final join unnecessarily materializes a large intermediate that is immediately re-aggregated for policy evaluation (\red{(1)} in \Cref{fig:rewriting}).

\twophase pushes the policy aggregate $\orange{\gamma_\Sigma}$ below the final join so it is applied to the relevant input relation (\red{(2)} in \Cref{fig:rewriting}).     

\begin{example}\label{ex:5}\it
Consider $Q=\gamma_{sum(y)}(T\Join S)$ from \Cref{fig:provstructs} and policy $P=(\{S\}, \gamma_{max(y)}(S) > 3, F()=\bot)$. 
The \logic approach has the following query structure, with the final joins in \red{red}:
$$\orange{\pi_{\lnot m}(\sigma_{m>3}(\gamma_{max(y)\to m}}(\gamma_{sum(y)\to s}(T\Join S)\red{\Join T \Join S})\orange{))}$$

In contrast, we can push the policy aggregation through the joins:
$$\orange{\pi_{\lnot m}(\sigma_{m>3}}(\gamma_{sum(y)\to s}(T\Join S)\red{\Join T \Join} \orange{\gamma_{max(y)\to m}}(\red{S})))\orange{)}$$

\end{example}

While this simple approach is effective and supports arbitrary aggregates, it can only push the policy aggregate through the final join.   If the query contains nested aggregates, then it cannot apply this push-down beyond the final join.

\subsection{$\onephase$ Rewriting}

Limitations of \twophase are that it requires re-scanning base relations to evaluate policy aggregates and that it cannot push enforcement through nested aggregations.  
$\onephase$ eliminates these extra scans by rewriting the {\it base query} so that policy aggregates are evaluated inline during normal query execution.  This algorithm is applicable for all semiring aggregates (see \Cref{sss:prov}).

Our observation is that if every aggregate in the policy predicate $B$ is a semiring, we can propagate the necessary semiring annotations through the query plan and ``piggyback'' computing partial policy aggregates during base query execution.  This sidesteps constructing provenance altogether.  
At a high level, $\onephase$ proceeds in two steps:

\begin{enumerate}[leftmargin=*]
\item \textbf{Top-down analysis.}  
Collect all semiring aggregates referenced in $B$. For each aggregate $\alpha$, identify the input relation $R$ it references and initialize the corresponding semiring annotation at $R$’s scan operator. 

\item \textbf{Bottom-up rewrite.}  
This phase propagates and computes partial aggregates throughout $Q$.   Joins in $Q$ combine annotations using $\cdot$, aggregations compute partial aggregates via $+$.  
\end{enumerate}

\subsubsection{Leaf Annotation.}
Let $B$ contain $n$ semiring aggregates.  
Each source relation referenced by $B$ is annotated with $n$ semiring elements, initialized according to the aggregate (e.g., $1$ for \code{count}, attribute value for \code{sum}, set containing tuple identifier for \code{distinct}).  
All other relations are annotated with the multiplicative identities $1$ of each semiring so that joins preserve correct multiplication semantics.   Redundant semirings do not need to be annotated twice.

\subsubsection{Operator Rules.}
Annotations are propagated as follows.
\textbf{Join} multiplies semiring annotations of joined tuples; 
\textbf{Projection} retains annotations unchanged,
\textbf{Aggregation} applies semiring addition to combine annotations within each group.


\begin{example}\it
Continuing \Cref{ex:5},
$\onephase$ augments the base aggregation to compute both aggregates simultaneously:
\[
\orange{\sigma_{m > 3}}
\left(
\gamma_{sum(s)\to s, \orange{max(m)\to m}}(T\Join_{id} \gamma_{id, sum(y)\to s, \orange{max(y)\to m}}(S))
\right).
\]
\noindent No additional joins with $S$ are required. The policy aggregate is evaluated inline with the base query aggregation.
\end{example}

\subsubsection{Handling Unions.}\label{sss:unions}

Union is the only disjunctive relational operator and requires special handling because it introduces alternative derivations from different sources.   To check whether a policy over source relation $R$ applies to an output tuple \code{o}, we need to check if $\code{o}$ is derived from $R$.  

We add an additional set semiring annotation $(2^D,\cup,\cup,\emptyset,\emptyset)$ that records the names of the contributing input relations, where $D$ is the set of relations in the database. Each input relation $R$ is annotated with $\{R\}$, and joins, projections, and unions propagate via $\cup$.  Thus each output tuple tracks its set of source relations. 

To check policy violations, we first check if the policy's sources are a subset of the annotation.   If not, the policy is skipped. While this construction is heavy-weight for unions, the next subsection describes its use for SQL disjunctive operators such as outer joins.

\subsubsection{Multi-Source} 
While the above procedure works for all relational algebra SPJUA queries and our core policy language, we note one special case where the policy specifies multiple sources that cannot always be pushed down.  

Consider the multi-source policy 
$P=(\{\code{T},$$\code{S}\},$ 
$\code{max(T.x+S.y)=15},$ 
$F()=\bot)$ over $Q=\code{T}\Join\gamma(\code{S})$.   
The expression  \code{T.x + S.y} does not distribute through \code{max} and thus cannot be pushed through the join.   In this case, we treat \code{(T.x,S.y)} as a source tuple, identify the lowest operator $op$ in the query plan whose output schema contains both attributes, and treat $op$ as a source relation. 
\onephase can be applied above $op$, and the subplan rooted at $op$ falls back to \logic or \twophase.  In contrast, \code{sum(T.x+S.y)=15} can be rewritten as \code{sum(T.x)+sum(S.y)=15} and be fully pushed down.






\subsection{Supporting TPC-H}
 To support all monotonic TPC-H queries, we now extend $\dfc$ policy evaluation and $\onephase$ rewriting beyond SPJUA relational algebra to additional SQL-92 operators.

\stitle{ORDER BY, LIMIT, OFFSET} do not change the derivation of individual output tuples, so no rewrites are needed.

\stitle{OUTER JOIN} introduces tuples with \code{NULL} values when one side has no matching contributor.  We use the same base relation annotations as for Union (\Cref{sss:unions}).  A tuple without one side of the join simply does not contain that input relation in its set annotation, and we can correctly check whether the policy applies to the tuple.

\stitle{SEMI JOIN, \code{EXISTS}, and \code{IN}} all short-circuit computation and may drop provenance multiplicities. Following PERM~\cite{glavic2009perm},  we rewrite semi-joins as \code{INNER JOIN}s to preserve correct derivation semantics. This handles correlated subqueries as well~\cite{glavic2009perm}.

\subsection{Supporting non-$\bot$ Resolutions}\label{ss:resolutions}
We now go beyond $\bot$ resolutions that filter violations to support transformation and relation-level resolutions.  

\subsubsection{Transformations.}\label{sss:transforms}
A resolution function $F()$ that transforms the violating tuple is challenging to handle because it {\it may} also return $\bot$.  Thus, the enforcement logic must take the output tuple's violation status and the output of $F()$ into account.   Let $Q'$ be the rewritten query's subplan immediately before the filter $\orange{\sigma_B}$ operator and $B$ be the policy's constraint.  We compute the following using common table expressions:
\begin{align*}
T_1 = &\pi_{*, B(*)\to b}(Q')\hspace{1.5em}&
T_2 = &\sigma_b(T_1)\\
T_3 = &\pi_{F(*)\to f}(\sigma_{\lnot b}(T_1))&
T_4 = &\pi_{f.*}(\sigma_{f\neq\bot}(T_3))\\
      &\code{return }T_2\cup T_4&&
\end{align*}
\noindent
$T_1$ computes whether each output tuple is a violation, and $T_2$ keeps all non-violating tuples.  $T_3$ evaluates the resolution function on each violation and stores the output as a tuple-valued attribute.  $T_4$ then removes $\bot$ tuples and promotes the tuple-value attribute into the output tuple (e.g., $((1,2))\to(1,2)$).   Finally, we return the union of the non-violating and transformed output tuples.
This naturally handles \kwcode{KILL} resolutions, which is simply a transformation resolution that throws an exception to terminate the query.  

\subsubsection{Relation-level.}
We model $F^{rel}$ as a user-defined table function whose output schema is schema compatible with the expected query output schema.   
Rather than filter violations using $\orange{\sigma_B}$, we instead annotate the output relation with whether each tuple is in violation or not.  Following the definitions in \Cref{sss:transforms}, we evaluate
$$F^{rel}(T_1)$$



\subsubsection{Non-blocking Resolutions.}
This work focuses on blocking resolution functions.   
In practice, we have implemented non-blocking functions to support e.g., user interfaces where users decide whether to evaluate, transform, or drop violations.   This requires custom operators in the query engine to asynchronously receive decisions from the UI, connection handling, and additional bookkeeping. Thus, we defer its presentation and evaluation to subsequent work.


\subsection{Optimizing Templated Policies} \label{ss:templated}

While it is unlikely for individual users to write policies from scratch, it is reasonable to predefine a templated policy that users customize for themselves.   For instance, \ex users may wish to customize the degree of k-anonymity they are comfortable with by specifying $K$:
\begin{FlowGuardExample}
SOURCE Receipts
CONSTRAINT COUNT(distinct Receipts.uid) > {K}
ON FAIL REMOVE
\end{FlowGuardExample}
\noindent Since the number of policies may scale to the number of users (in the millions), we wish to support a useful class of templates that can be enforced in sublinear time.  
In this paper, we describe a class of single-attribute comparative predicates that satisfy this goal:
\begin{verbatim}    {agg}({expr}) {op} {K} \end{verbatim}
Each bracketed term can be parameterized.  \code{agg} is an aggregation function, \code{expr} is a single-attribute expression (e.g., \code{risk+0.5}), \code{op} is a comparator (e.g., >, <, =), and \code{K} is a numeric threshold.    We focus on this class because it is amenable to algebraic optimization.  

Let us first assume \code{agg(expr)} is predefined and fixed, and will relax this assumption below.   Under conjunction, comparison operators fall into three algebraic classes.
\begin{itemize}[leftmargin=*]
    \item \stitle{Monotone threshold operators} ($>, \ge, <, \le$) collapse to a single dominating threshold: lower bounds ($>, \ge$) reduce to the maximum $K$, and upper bounds ($<, \le$) to the minimum $K$.
    \item \stitle{Exact-match operators} ($=$) are either redundant (if thresholds agree) or unsatisfiable.
    \item\stitle{Exclusion operators} ($\ne$) collapse to one \code{NOT IN} over all $K$.
\end{itemize}

\noindent To parameterize \code{\{agg\}(\{expr\})} as well, we apply the above strategy for every unique aggregation expression, and use algebraic manipulation to simplify e.g., \code{sum(x+1)=K} to \code{sum(x)+count(x)=K}.
In practice, we maintain a data structure that updates each time the template is used to create a new policy.  
We then generate the conjunction of all distinct predicate clauses, and pass it to the enforcement engine.

More general templated policies that contain disjunctions would be useful for more expressive individualized policies like {\it ``if my data is used, then it should be mixed with K other users.''}  However, it is easy to reduce such policy enforcement into NP-hard problems, and we leave identifying scalable but expressive template classes and pragmatic approximations as promising future directions.

\subsection{Optimizing Self Joins}\label{ss:selfjoins}
Queries containing self joins are challenging to enforce because they require a factorial number of policy evaluations.  
Formally, suppose a query contains $n$ instances of relation $R$, and a policy’s \code{SOURCE} clause references $m$ aliases of $R$. A naive enforcement evaluates the policy for every assignment of policy aliases to query aliases---$_nP_m$ (permutations) policy applications. For example, applying the following policy to $Q=R_1\Join R_2\Join R_3$ needs $_3P_2 = 6$ policy evaluations. 
 \begin{FlowGuardExample}
SOURCE Receipts R1, Receipts R2
CONSTRAINT R1.year = R2.year
ON FAIL REMOVE
\end{FlowGuardExample}
\noindent  We identify a common case where enforcement scales linearly. A policy is \emph{symmetric} over aliases $R_1,\dots,R_m$ if it is invariant under permutation of those aliases. Concretely, symmetric policies consist of i) unary predicates $u(T_i)$ applied uniformly to each alias and ii) symmetric binary predicates $b(R_i,R_j)$ that are invariant under swapping arguments (e.g., equality of aggregates).

Symmetric policies suffice to evaluate the predicate once over the set of all $n$ query aliases. Rather than enumerating $_nP_m$ permutations, we rewrite the predicate to apply the unary clauses to each alias and enforce binary predicates using $n-1$ comparisons against a representative alias. This reduces enforcement cost from factorial to linear in $n$.
We show this in \Cref{sss:selfjoin}.

\section{Evaluation} \label{sec:eval}

We now evaluate the enforcement methods from \Cref{sec:rewriting}.  We first evaluate TPC-H, then consider the trade-off between \onephase and \twophase, then study \onephase in depth, and finally show how policies can enforce process compliance described in \Cref{ss:usecases}.


\subsection{Setup}
We compare $\logic$, $\phys$, $\onephase$, $\twophase$ enforcement mechanisms against the query without enforcement (\code{No Policy}). 
In practice, we expect most queries are not policy-violating.  Further, policies that don't prune data flows should incur the highest overhead, so all evaluated policies do not drop rows.

By default, we run on DuckDB v1.3.0.  \phys, use the SmokedDuck extension\footnote{\url{https://github.com/haneensa/lineage}}.  
\Cref{sec:multi-engine} compares five major DBMS engines that vary in architecture: DuckDB \cite{raasveldt2019duckdb}, Umbra \cite{neumann2020umbra}, PostgreSQL \cite{stonebraker1986postgres}, DataFusion \cite{lamb2024datafusion}, and SQLServer \cite{microsoft2026sqlserver}.   
All experiments are run on a 2024 MacBook Air\footnote{DuckDB and Datafusion were run in-process, Umbra and Postgres were run in Docker.} (8-core M3, 16GB RAM) except for SQL Server, which is run on AWS RDS (4 vCPUs, 16GB RAM). We report averages over 5 runs after a warmup run.



\subsection{TPC-H}
We evaluate all monotonic TPC-H queries.  First, we vary scale factor on DuckDB and analyze runtime overhead.  Second, we run on all DBMS engines. All monotonic queries scan \code{lineitem}, so we enforce the following policy:
\begin{FlowGuardExample}
SOURCE lineitem
CONSTRAINT max(lineitem.l_quantity) >= 1
ON FAIL REMOVE
\end{FlowGuardExample}

\subsubsection{Database Scaling}

\begin{figure}
    \centering
    \includegraphics[width=\linewidth]{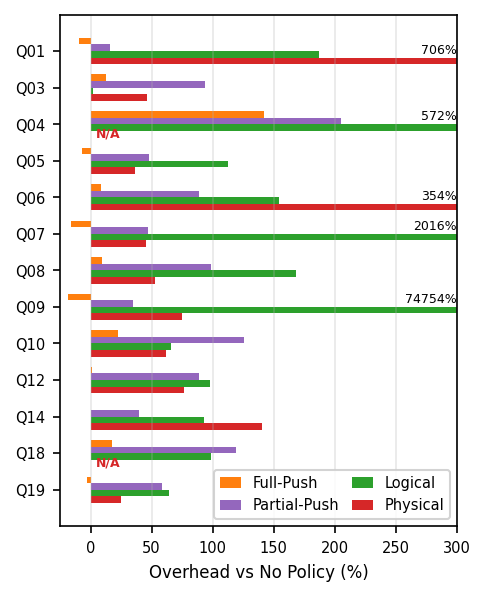}
    \caption{Relative overhead of policy enforcement methods (TPC-H SF=10). $\phys$ cannot be run on Q4,18. $\onephase$ achieves $\sim 0\%$ overhead on all queries except Q4. }
    \label{fig:tpch_sf10}
\end{figure}

We find that relative overheads remain consistent across SF=1 and 10, so report SF=10 (\Cref{fig:tpch_sf10}).  \phys does not run Q4,18 because they contain semi joins that are rewritten as inner joins with an inner aggregation and the current SmokedDuck implementation does not support nested aggregation.


\onephase is consistently the fastest and does not exceed single-digit relative overheads except for Q4.  This is because Q4 contains an semi join (\code{EXISTS (SELECT * FROM lineitem WHERE ...}) that all approaches must rewrite into an inner join. $\logic$ slows down by >500\% because annotated output $O$ is large.


Considering other approaches, $\logic$ has particularly poor performance (2,000\% - 75,000\% slowdown) when $O$ is very large because there are many derivations (Q7,9). We see that DuckDB does not rewrite $\logic$ into $\twophase$, otherwise $\logic$ would be similarly performant. $\phys$ performs poorly (300\% - 700\% slowdown) when there are few outputs with many contributors (Q1,6) since the join it makes between lineage and base tables is large.

\subsubsection{DBMS Engines} \label{sec:multi-engine}

\begin{figure}
    \centering
    \includegraphics[width=\linewidth]{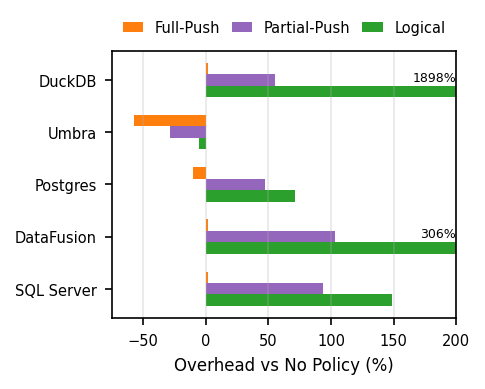}
    \caption{Relative overhead to enforce 1 policy across 5 major DBMS engines. $\onephase$ incurs nearly zero overhead, and \twophase is considerably faster than \logic.  Surprisingly, policy enforcement is faster than \code{No Policy} on Umbra for all approaches.}
    \label{fig:multi_db}
\end{figure}

\Cref{fig:multi_db} reports the relative overhead of enforcing the same \code{lineitem} policy for 5 popular DBMS engines, averaged across the TPC-H queries on SF=1. We don't include $\phys$ because it is built into to DuckDB. 
Across the board, $\onephase$ is dominant --- it is on average faster than the base query across engines. Surprisingly, all policy evaluation performance improves in Umbra \cite{neumann2020umbra}. There is a gap (20\% to >1000\%) between $\twophase$ and $\logic$ highlighting that the aggregate push down that $\twophase$ performs is not consistently found by any engine's optimizer.

\subsubsection{Why is \onephase and \twophase So Fast?} \label{sss:sofast}

We now use Q9 to analyze the reasons why \onephase and \twophase is faster than the baselines.  We report \logic since \phys exhibits similar patterns.    We find that the primary cost comes from aggregation operators, which \logic must translate into provenance propagation then a post-processing join operation.  This causes huge intermediate result sizes that contribute to the overhead.   

\begin{figure}
    \centering
    \includegraphics[width=.8\linewidth]{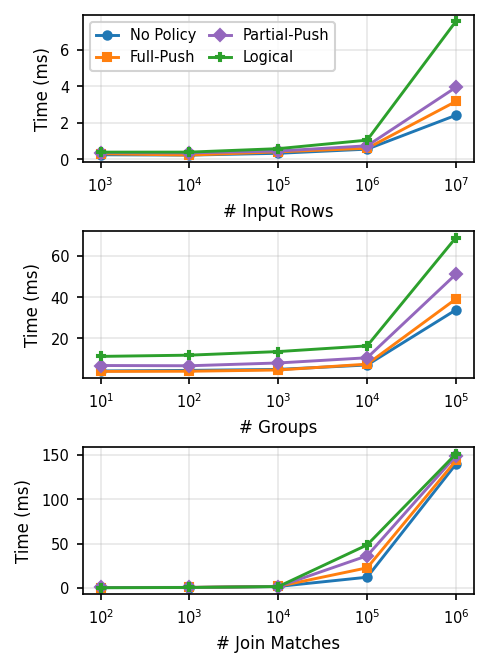}
    \caption{We consider single operator base queries. Simple aggregation (Top) across various sizes. Group by (Middle) performance varying the number of distinct groups. Join (Bottom) performance varies the number of join matches.}
    \label{fig:micro}
\end{figure}

To illustrate this, we evaluate two aggregation-only queries: a \code{Simple} aggregation with no grouping attributes where we vary the input cardinality from $1K-10M$ (\Cref{fig:micro} Top Left), and a \code{Grouping} aggregation with fixed $1M$ input cardinality and vary the number of groups from $10-100K$ (\Cref{fig:micro} Top Right). Recall that $\logic$ rewrites base query aggregations to be followed by a join with the base relation -- this expands $O$ to include a separate row for each derivation. \onephase avoids re-scanning and joining the table to evaluate the policy. Both \onephase and \twophase avoid building large intermediates by pushing the policy aggregation into the second scan before joining.

To verify that non-aggregation operators are not the bottleneck, \Cref{fig:micro} (Bottom) evaluates a join query; we fix the input cardinality of first table at $10M$ and vary the number of tuples in the second table from $100-1M$. Each row has a single join match. We see that the overhead of all methods are almost indistinguishable (\onephase is slightly faster) from \code{No Policy}.




\subsection{Comparing $\onephase$ and $\twophase$} \label{sec:comp}



Across the other experiments, $\onephase$ is either equivalent to or better than $\twophase$. But $\twophase$ is dominant when adding attributes to intermediates results in poor cache locality within the base query. This experiment highlights that phenomena.

The base query considers 128 source relation attributes joined with a second table. The number of distinct attributes considered in the policy varies by powers of 2 from 2 to 512. The base query join fanout varies by powers of 2 from 2 to 64 to scale intermediate size.

\begin{figure}
    \centering
    \includegraphics[width=\linewidth]{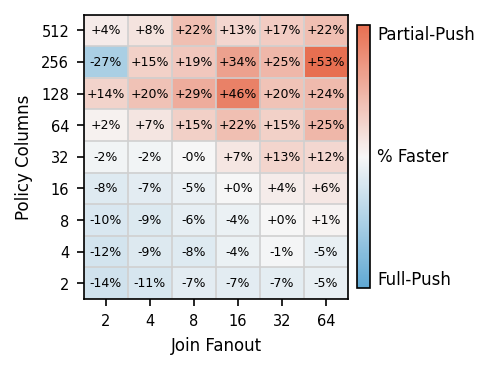}
    \caption{Highlighting where (\red{Red}) \red{$\twophase$ outperforms $\onephase$}, or (\blue{Blue}) \blue{\onephase outperforms \twophase.}}
    \label{fig:full_vs_partial}
\end{figure}

\Cref{fig:full_vs_partial} shows that $\twophase$ is dominant (Red) when intermediates are larger and there are more distinct policy columns.


\subsection{Studying \onephase in Depth}

The preceding experiments suggest that \onephase is, in general, the lowest overhead enforcement method.  We now study the method in depth to understand how it scales under policy complexity and can be optimized (\Cref{ss:templated}, \Cref{ss:selfjoins}).

\begin{figure}
    \centering
    \includegraphics[width=\linewidth]{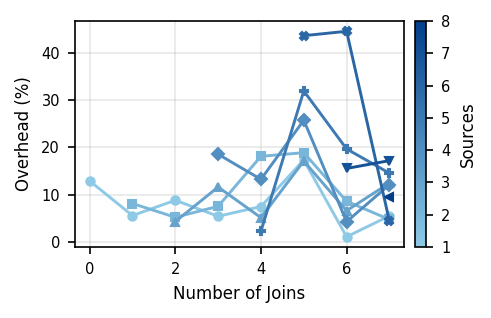}
    \caption{Scaling the number of sources in a $\onephase$ policy on a TPC-H Q1 variant. A policy with $n$ sources is applied if join count is $\geq n$.}
    \label{fig:many_source}
\end{figure}

\subsubsection{Policy Complexity: Number of Sources}
\Cref{sec:comp} evaluates the constraint complexity as the number of expressions scales up.   In this experiment, we study how \onephase scales as the number of sources in one policy increases.  

To do so, we scale both the number of sources in the policy and in the query from 1 to 8.  For the query, we start with a modified Q1 --- $\gamma(\sigma(\code{lineitem}))$ --- that starts as-is only reading \code{lineitem}, then incrementally add $1-7$ joins with other TPC-H relations by foreign key relationships. The policy sources vary between 1 (just \code{lineitem}) to 8.   Note that the policy only applies to queries with the same or more source relations.
\Cref{fig:many_source} reports that compared to \code{No Policy}, the relative overhead is not impacted by adding sources.

\subsubsection{Number of Policies}
This experiment highlights the impact of many policies. As described in \Cref{ss:templated}, millions of templated policies may be common as many individuals provide policies regarding their data. This experiment considers 1-1M policies over TPC-H Q1 at scale factor 1. We consider $\onephase$ compared with $\onephase$ \code{Optimized} which applies \Cref{ss:templated}'s optimization.

\begin{figure}
    \centering
    \includegraphics[width=\linewidth]{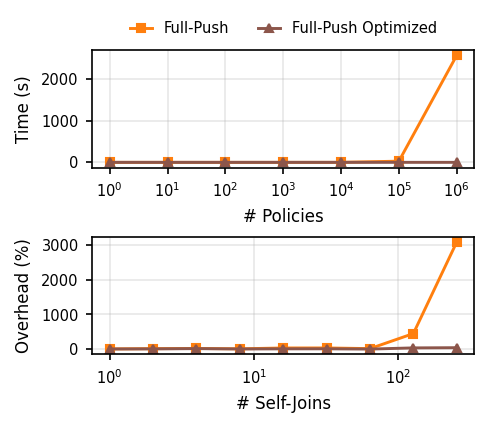}
    \caption{Performance of TPC-H Q1 across many templated policies (Left) and a variant of Q1 across many self joins with a multi source policy with multiple \kwcode{SOURCE} references to the self-joined table (Right). These figures highlight optimizations from \Cref{ss:templated} and \Cref{ss:selfjoins}.}
    \label{fig:many_policies_and_self_joins}
\end{figure}

The results in \Cref{fig:many_policies_and_self_joins} (Left) show the performance of the optimization -- it identifies and runs a single dominant policy, so policy evaluation is constant even as the number of policies explodes.

\subsubsection{Many Self Joins} \label{sss:selfjoin}

Policies with multiple sources that refer to the same relation $m$ times are applied to queries that self join $n$ times across every permutation, so $_nP_m$. \Cref{ss:selfjoins} optimizes the useful class of symmetric policies so the policy is applied $O(n)$ times instead of $O(n!)$ times. We evaluate over a symmetric policy with 2 references to the \code{lineitem} table and scale an augmented version of TPC-H Q1 that self-joins 1-1k times.

\Cref{fig:many_policies_and_self_joins} (Right) highlights the optimization -- at 1k self-joins, the policy is naively applied >1m times, but the optimized approach only adds 1k predicates.

\subsection{Application Workload}


We highlight an application workload designed to represent Business Processes (\Cref{sss:processes}). We encode the state machine

{\centering\begin{tikzpicture}[->, >=stealth, node distance=2cm]
  \node[state] (A) {$A$};
  \node[state, right of=A] (B) {$B$};
  \node[state, right of=B] (C) {$C$};

  \path
    (A) edge (B)
    (B) edge (C)
    (B) edge[bend left] (A);
\end{tikzpicture}}

\noindent into the following policy:
\begin{FlowGuardExample}
SOURCE T AS T1
SINK T AS T2
CONSTRAINT T1.id = T2.id AND
    CASE WHEN T1.state = 'A' THEN T2.state = 'B'
    WHEN T1.state = 'B' THEN T2.state IN ('A', 'C')
    WHEN T1.state = 'C' THEN false END
ON FAIL REMOVE
\end{FlowGuardExample}

\noindent $T(id, state)$ is initialized with 1000 items all where $state = A$. We then run 1000 $UPDATE$ statements on $T$ where 70\% are valid state transitions randomly sampled from the state machine, and 30\% are invalid (e.g., $C\to B$). We compare \code{No Policy}, $\onephase$, and \code{GPT-5.2}. For \code{GPT-5.2}, the prompt describes the state machine and asks if a given UPDATE query is allowed.

\begin{figure}
    \centering
    \includegraphics[width=\linewidth]{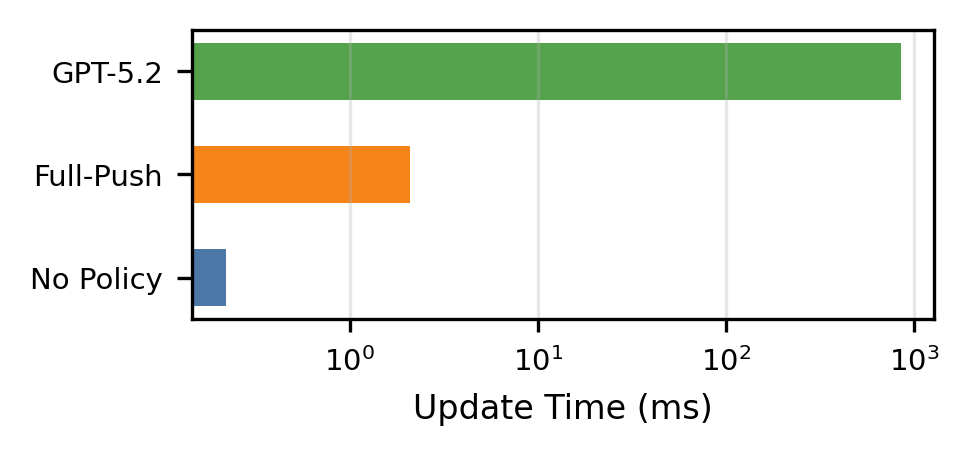}
    \caption{We compare \code{No Policy}, $\onephase$, and \code{GPT-5.2} to maintain a state machine. \code{No Policy} allows illegal state transitions. $\onephase$ and \code{GPT-5.2} are both correct, and $\onephase$ is 2 orders of magnitude faster.}
    \label{fig:state_transition}
\end{figure}

Both $\onephase$ and \code{GPT-5.2} correctly stops all invalid transitions and allows all valid transitions. $\onephase$ is 2 orders of magnitude faster than \code{GPT-5.2}.  \onephase's overhead compared to \code{No Policy} is because the \code{UPDATE} query is trivial; the cost mainly comes from needing to evaluate the CASE WHEN in the \kwcode{CONSTRAINT}.  

\section{Related Work} \label{sec:related_work}

Provenance has been explored for access control \cite{park2012pbac,upadhyaya2015datalawyer} and access control has been applied to provenance \cite{braun2008securingprov,cadenhead2011paclang}. The work most similar to ours is DataLawyer \cite{upadhyaya2015datalawyer} which supports compliance- and regulation-motivated policies like ours, as well as policies that refer to the provenance of other queries. DataLawyer's expressivity comes at a cost: when capturing provenance it imposes a >100\%  slowdown (provenance tax) for all queries.

Several recent papers \cite{debenedetti2025camel,tsai2025conseca,li2025safeflow,costa2025fides,beurerkellner2025llmdesignpatterns} consider deterministic agent safety to defend against Prompt Injection. Ultimately, they propose coarse-grained guardrails restricting agent access to tools or data. Because they don't track data flows between the agent, sensitive data, and exfiltrating tools, they disallow safe actions.

Semantic Integrity Constraints~\cite{lee2025semantic} provide grounding by ensuring that LLM outputs are semantically consistent with their inputs, enforced via substring containment (extractive cases) or LLM-based judges (abstractive cases). In contrast, our notion of grounding constrains the provenance relationship between source and sink tuples, enabling deterministic guarantees that outputs are derived from specific inputs rather than merely judged consistent.

Factorized query processing systems (LMFAO~\cite{schleich2019lmfao}, SDQL~\cite{shaikhha2021sdql}, Morpheus~\cite{chen2017morpheus}) optimize queries by pushing down aggregates and factorizing computation similar to $\onephase$. Our contribution applies factorization principles to a new setting: efficiently evaluating boolean predicates over provenance monomials without materializing provenance.

\section{Conclusion} \label{sec:conclusion}

Agents increasingly issue queries, orchestrate pipelines, and automate analyses.  While queries must reflect the user's semantic goals, this alone is not sufficient.  We must guarantee {\it data safety}---that all data flows comply with regulatory, privacy, business, and safety constraints. We introduced Data Flow Control (\dfc), the first framework to declaratively specify and enforce safety policies over tuple-level data flows inside a DBMS.  We formalized a policy language $\lang$ that is expressive, optimizer-invariant, and fast to enforce. We further designed a lightweight enforcement engine $\sys$ that rewrites enforcement into query execution and avoids the cost of provenance materialization.   We show near-zero overhead across DuckDB~\cite{raasveldt2019duckdb}, Umbra~\cite{neumann2020umbra}, PostgreSQL~\cite{stonebraker1986postgres}, DataFusion~\cite{lamb2024datafusion}, and SQLServer~\cite{microsoft2026sqlserver}; conditions that scale enforcement to millions of policies; and provide guarantees that current frontier models do not provide (and at orders of magnitude higher cost and latency).  
We believe that starting from a fast and scalable design removes concerns about cost from the table, and is a crucial step toward safe-by-default data infrastructure.  Our future work will move toward larger-scale policy enforcement across multiple queries and general purpose agentic workflows.

\bibliographystyle{ACM-Reference-Format}
\bibliography{sample}

\end{document}